\documentclass[12pt]{iopart}
\usepackage{iopams}
\usepackage{graphicx}
\input amssym.def \input amssym

\begin{document}

\title[]{Entangling gates in even Euclidean lattices such as the Leech lattice}

\author{Michel Planat$^{1}$}

\address{ $^1$ Institut FEMTO-ST, CNRS, 32 Avenue de l'Observatoire, F-25044 Besan\c con, France.}

\begin{abstract}
The group of automorphisms of Euclidean (embedded in $\mathbb{R}^n$) dense lattices such as the root lattices $D_4$ and $E_8$, the Barnes-Wall lattice $\mbox{BW}_{16}$, the unimodular lattice $D_{12}^+$ and the Leech lattice $\Lambda_{24}$ may be generated by entangled quantum gates of the corresponding dimension. These (real) gates/lattices are useful for quantum error correction: for instance, the two and four-qubit real Clifford groups are the automorphism groups of the lattices $D_4$ and $\mbox{BW}_{16}$, respectively, and the three-qubit real Clifford group is maximal in the Weyl group $W(E_8)$.
Technically, the automorphism group $\mbox{Aut}(\Lambda)$ of the lattice $\Lambda$ is the set of orthogonal matrices $B$ such that, following the conjugation action by the generating matrix of the lattice, the output matrix is unimodular (of determinant $\pm 1$, with integer entries). When the degree $n$ is equal to the number of basis elements of $\Lambda$, then $Aut(\Lambda)$ also acts on basis vectors and is generated with matrices $B$ such that the sum of squared entries in a row is one, i.e. $B$ may be seen as a quantum gate.
For the dense lattices listed above, maximal multipartite entanglement arises. In particular, one finds a balanced tripartite entanglement in $E_8$ (the two- and three- tangles have equal magnitude $1/4$) and a GHZ type entanglement in BW$_{16}$. In this paper, we also investigate the entangled gates from $D_{12}^+$ and $\Lambda_{24}$, by seeing them as systems coupling a qutrit to two- and three-qubits, respectively.
Apart from quantum computing, the work may be related to particle physics in the spirit of \cite{PLS2010}.

\end{abstract}

\pacs{03.65 Ud, 03.67.Pp, 02.20.-a}

\normalsize
\section{Introduction}
 
Entanglement is at the heart of many paradoxes of quantum mechanics and is also a useful ressource in quantum information processing. In particular, entangled states play a key role in quantum error correction  and in quantum computational speedup. The characterization of entanglement of a general multipartite state remains a mathematical challenge and there exist several inequivalent measures. A comprehensive account of the most important mathematical concepts used for quantifying and manipulating entangled states may be found in \cite{Horodecki}.

Some properties of entangled states may also be approached from differential geometry (two- and three-qubit Hilbert space geometry  corresponds to Hopf fibrations, that are entanglement sensitive \cite{Hopf}), from topological spaces (for instance the $3$-qubit GHZ state corresponds to a borromean ring \cite{Aravind}), from finite geometries (by using the hyperplanes of the relevant generalized quadrangles \cite{PauliGraphs} or in the black-hole analogy \cite{Levay}. Besides, group theoretical concepts have been repetitively put forward for describing classical \cite{Nebe} and quantum error correction \cite{Calderbank}. In this paper, we exploit the unnoticed but organic link between real entangling gates of quantum computation and the group of automorphisms of some highly symmetric even Euclidean lattices. 

Let us define an $n$-dimensional Euclidean lattice $\Lambda$ as a discrete additive subgroup of the real vector space $\mathbb{R}^n$, endowed with the standard Euclidean product, and spanned by a generator matrix $M$ with rows in $\mathbb{R}^n$.  The automorphism group $\mbox{Aut}(\Lambda)$ is the set of orthogonal matrices $B$ such that under the conjugation action by the generating matrix $M$, one gets a unimodular matrix $U=MBM^{-1}$, that is (i) $\mbox{det}~ U= \pm 1$ and (ii) $U$ is an integer matrix (see \cite{Conway}, p. 90). When the degree $n$ is equal to the number of basis elements of $\Lambda$, then $Aut(\Lambda)$ also acts on basis vectors and is generated with matrices $B$ such that the sum of squared entries in a row is one, i.e. $B$ may be seen as a quantum gate.

It will be shown that for a suitable choice of even Euclidean lattices $\Lambda$ such as $\mathbb{Z}^n$ lattices, the root lattices $D_4$ and $E_8$, the Barnes-Wall lattice $\Lambda_{16}$, the lattice $D_{12}^+$ and the Leech lattice  $\Lambda_{24}$, one observes that the orthogonal group $\mbox{Aut}(\Lambda)$ is a group of entangling {\it quantum gates}. It can be considered as  acting on a two-, three-, four-, a two-qubit/qutrit and a three-qubit/qutrit system, respectively. In retrospect, such lattices bear out our recent work relating $D_4$ and $E_8$ Weyl groups to entangled states arising from the joint basis of specific mutually commuting set of generalized Pauli observables \cite{PLS2010,PlanatCPT}.

\section{Entanglement arising from $\mathbb{Z}^n$ lattices}

To illustrate our topic, let us start with the $\mathbb{Z}^n$ lattice whose lattice points are the integers and whose basis $M$ is the  identity matrix. The elementary cell is the $n$-dimensional hypercube and the automorphism group is a wreath product \cite{Harrary}
\begin{equation}
\mbox{Aut}(\mathbb{Z}^n) \cong \mathbb{Z}_2 \wr S_n=\mathbb{Z}_2^n \rtimes S_n, 
\end{equation}
of order $2^n n!$, where the symbol $\rtimes$ denotes the semidirect product and the wreath product $\wr$ corresponds to a permutation action of the symmetric group $S_n$ on the $n$ copies of the two-letter group $\mathbb{Z}_2$. 

It is straightforward to check that the automorphism group $\mbox{Aut}(\mathbb{Z}^4)$ of the four-dimensional lattice contains the gate CNOT=$\left(\begin{array}{cccc} 1 & 0 & 0 & 0 \\0 & 1 & 0 & 0 \\ 0 & 0 & 0 & 1 \\0 & 0 & 1 & 0\\ \end{array}\right)$, that is well known to create two-qubit entanglement. More precisely, taking the input qubit states as $\left|0\right\rangle$ and $\left|\psi\right\rangle=a\left|0\right\rangle+b\left|1\right\rangle$,  the action CNOT$(\left|0\right\rangle\left|\psi\right\rangle)$ ends up into the non-separable Bell state $a\left|00\right\rangle+b\left|11\right\rangle$. As it is well known, the CNOT gate is a primitive of universal quantum computing .

It has been found \cite{PLS2010} that the entanglement arising from the two-qubit Pauli group is encapsulated in the following  Mermin's square
\begin{eqnarray}
&\sigma_x \otimes \sigma_x~~\sigma_y \otimes \sigma_y~~\sigma_z \otimes \sigma_z \nonumber \\
&\sigma_y \otimes \sigma_z~~\sigma_x \otimes \sigma_z~~\sigma_x \otimes \sigma_y \nonumber \\
&\sigma_z \otimes \sigma_y~~\sigma_z \otimes \sigma_x~~\sigma_y \otimes \sigma_x, \nonumber \\
\label{MSquare}
\end{eqnarray}
where $\sigma_x$, $\sigma_y$ and $\sigma_z$ are the ordinary Pauli spin matrices. Every row and every colum in the above square consists of mutually commuting operators sharing a basis of entangled states. The Mermin's square may be used for a proof of the Kochen-Specker theorem in a four-dimensional space by observing that the algebraic relations for the eigenvalues $\pm 1$ of the observables contradicts the one for the eigenstates \cite{PLS2010,Mermin}.

Let us denote $r=\{\sigma_x \otimes \sigma_x,\sigma_y \otimes \sigma_y,\sigma_z \otimes \sigma_z\}$ and $s=\{\sigma_x \otimes \sigma_z, \sigma_z \otimes \sigma_x,\sigma_y \otimes \sigma_y\}$ the first row and the r.h.s. column  in (\ref{MSquare}), the only ones that possess real entries. One may use the common eigenstate basis $M_r=\{(1,0,0,1),(1,0,0,-1),(0,1,1,0),(0,1,-1,0)\}$ of the set $r$ for generating a lattice isometric to $\mathbb{Z}^4$. The natural action of the automorphism group on the basis $M_r$ reads \footnote{The reader may use the following commands in Magma \cite{Magma} to check the calculations: \lq\lq Mr:=Matrix(4,4,[1,0,0,1,  1,0,0,-1,  0,1,1,0,  0,1,-1,0]);
L:=LatticeWithBasis(Mr);
aut:=AutomorphismGroup(L:NaturalAction);
L,aut;"}.
\small
\begin{equation}
\mbox{Aut}(\mathbb{Z}^4)=\left\langle S=\frac{1}{2}\left(\begin{array}{cccc} 1 & -1 & 1 & 1 \\1 & 1 & 1 & -1 \\ 1 & -1 & -1 & -1 \\1 & 1 & -1 & 1\\ \end{array}\right),
S'=\frac{1}{2}\left(\begin{array}{cccc} 1 & -1 & 1 & -1 \\-1 & 1 & 1 & -1 \\ 1 & 1 & 1 & 1 \\-1 & -1 & 1 & 1\\ \end{array}\right) \right\rangle,
\label{autZ4}
\end{equation}
\normalsize
It is noticeable that the rows of the generators $S$ and $S'$ of $\mbox{Aut}(\Lambda)$ encode the eigenstates of the triples of observables $s$ and $\{\sigma_x \otimes 1_2,1_2 \otimes \sigma_x, \sigma_x \otimes \sigma_x\}$, respectively (a partial list of the eigenstates, associated to the maximal sets of mutually commuting operators in the two-qubit Pauli group, may be found in \cite{PlanatSigma}).

Then, following a similar reasoning, we construct the  $\mathbb{Z}^8$ lattice by means of a generating basis built from the eigenstates of the following triple of three-qubit observables 
\begin{equation}
s_3=\sigma_z \otimes\{\sigma_x \otimes \sigma_z, \sigma_z \otimes \sigma_x,\sigma_y \otimes \sigma_y\},
\label{triples3}
\end{equation}
that follows from $s$ by applying the left tensor product $\sigma_z$. The basis we select is the matrix $S_3$ first introduced in \cite{PLS2010} (Eq. (9)]. As for the two-qubit case above, the natural action of the automorphism group on the basis is obtained as

\footnotesize
\begin{eqnarray}
&\mbox{Aut}(\mathbb{Z}^8)=\left\langle g_1,g_2 \right \rangle, \nonumber \\
 &\mbox{with}~~ g_1=\frac{1}{2}\left(\begin{array}{cccccccc} . & . & 1 & 1 & . & 1& .& 1 \\  & . & 1 & 1 & . & -1& .& -1 \\-1 & 1 &. &. &1 &.& -1& . \\1 & -1 &. &. &1 &.& -1& . \\ 
1 & 1 &. &. &. &1&.& -1 \\1 & 1 &. &. &. &-1&.& 1 \\  . & . & -1 & 1 &1 & .& 1& . \\    . & . &1 & -1 &1 & .& 1& . \\    \end{array}\right).
\end{eqnarray}
\normalsize
The matrix $g_2$ is not made explicit but it has a similar form and properties as $g_1$. Let us now investigate the type of entanglement contained in the $3$-qubit generators/gates $g_1$ and $g_2$. We single out the state encoded by the fourth row of $g_1$
\begin{equation}
\left|\psi\right\rangle=\frac{1}{2}(\left|000\right\rangle-\left|001\right\rangle+\left|100\right\rangle-\left|110\right\rangle).
\label{state}
\end{equation}
A quantitative measure of tripartite entanglement is the $3$-tangle \cite{FourTangle,Coffman}
\begin{equation}
\tau^{(3)}=4\left|(T^{001}-T^{000})^2 -4P_{B_1}^{00}P_{B_0}^{00}\right|,
\label{three}
\end{equation}
which contains the four determinants \\$T^{000}=\mbox{det}\left(\begin{array}{cc} a_{000} & a_{011}  \\a_{100} & a_{111} \\ \end{array}\right)$, $T^{001}=\mbox{det}\left(\begin{array}{cc} a_{001} & a_{010}  \\a_{101} & a_{110} \\ \end{array}\right)$,\\ $P_{B_0}^{00}=\mbox{det}\left(\begin{array}{cc} a_{000} & a_{001}  \\a_{100} & a_{101} \\ \end{array}\right)$ and $P_{B_1}^{00}=\mbox{det}\left(\begin{array}{cc} a_{010} & a_{011}  \\a_{110} & a_{111} \\ \end{array}\right)$.

To quantify bipartite entanglement the strategy is as follows \cite{Coffman}. Take the density matrix $\left|\psi\right\rangle \left\langle \psi\right|$ of the $3$-qubit system and trace out over the bipartite subsystems to obtain a reduced density matrix $\rho$. A measure of two-qubit entanglement is the tangle $\tau=C^2$, with the concurrence $C(\rho)=\mbox{max}\{0,\sqrt{\lambda_1}-\sqrt{\lambda_2}-\sqrt{\lambda_3}-\sqrt{\lambda_4}\}$, where the $\lambda_i$ are non-negative eigenvalues of the product $\rho \tilde{\rho}$ arranged in decreasing order, $\tilde{\rho}=(\sigma_y \otimes \sigma_y)\rho^{\ast}(\sigma_y \otimes \sigma_y)$ is the spin-flipped density matrix and $\ast$ denotes the complex conjugate.

Two important families are the GHZ family with $\tau^{(3)}=1$ and all $2$-tangles $\tau$ vanishing, and the $W$ family with $\tau^{(3)}=0$ and all $2$-tangles equal to 1.  Both families are unequivalent under SLOCC transformations \cite{Coffman}. 

The $3$-tangle of the state (\ref{state}) is $\tau^{(3)}=\frac{1}{4}$ and the reduced density matrices for subsystems $A-B$, $A-C$ and $B-C$ are
$\rho_{AB}=\frac{1}{4}\left(\begin{array}{cccc} 2 & . & 1 &- 1 \\. & . & . & . \\ 1 & . & 1 & -1 \\-1 & . & -1 & 1\\ \end{array}\right)$, $\rho_{AC}=\frac{1}{4}\left(\begin{array}{cccc} 1 & -1 & 1 &. \\-1 & 1 & -1 & . \\ 1 & -1 & 2 & . \\.& . & . & .\\ \end{array}\right)$ and $\rho_{BC}=\frac{1}{4}\left(\begin{array}{cccc} 2 & -1 & -1 &. \\-1 &1 & . & . \\ -1 & . & 1 & . \\. & . & . & .\\ \end{array}\right)$. The set of eigenvalues of $\rho \tilde{\rho}$ associated to subsystems $A-B$ and $A-C$ are $\{\frac{1}{16}(3+2\sqrt{2}),\frac{1}{16}(3-2\sqrt{2}),0,0\}$ leading to the tangles $\tau_{AB}=\tau_{AC}=\frac{1}{4}$ and the set of eigenvalues of $\rho \tilde{\rho}$ associated to the subsystem $B-C$ is $\{\frac{1}{16},\frac{1}{16},0,0\}$ leading to a vanishing tangle $\tau_{BC}$. 

All states encoded by rows of $g_1$ and $g_2$ behaves as state (\ref{state}). We call a tripartite entanglement characterized by the tangles $\tau^{(3)}=\tau_{AB}=\tau_{AC}=\frac{1}{4}$ and $\tau_{BC}=0$ an {\it incomplete balanced entanglement}. Fully balanced tripartite entanglement is found in \cite{PLS2010,PlanatCPT} and in the next section below devoted to the $E_8$ lattice and other dense lattices.

\section{Entanglement arising from Barnes-Wall lattices}

The family of Barnes-Wall lattices $L_n$, of even dimension $2^n$, is related to self-dual codes and the real Clifford group. The first four members of the family, for $n=1$ to $4$, are isomorphic to the densest known lattices of the corresponding dimension, i.e. the root lattices $A_2$, $D_4$, $E_8$ and the Barnes-Wall lattice $\mbox{BW}_{16}$. 

The automorphism group of $L_n$ (in dimension $2^n$, $n \ne 3$) corresponds to the real Clifford group $\mathcal{C}_n^{+}$ (defined as the normalizer in the general orthogonal group $O(2^n)$ of the generalized Pauli group on $n$ real qubits: the notation comes from \cite{PlanatCPT}). For $n=3$ qubits, $L_3$ is $E_8$ and $\mathcal{C}_3^+$ is the second largest maximal subgroup of $W'(E_8)$, the derived subgroup of the Weyl group $W(E_8)$. In the sequel, we concentrate on the entanglement arising from the lattices embodying two, three and four qubits, respectively. 

\subsection*{Two-qubit entanglement and the lattice $D_4$}

Among the several representations of the lattice $D_4$, we select the one with basis $M_r=\{(1,1,0,0),(1,-1,0,0),(0,1,-1,0),(0,0,1,-1)\}$, that has the automorphism group  
\small
\begin{equation}
\mbox{Aut}(D_4)=\left\langle \frac{1}{2}\left(\begin{array}{cccc} 1 & -1 & 1 & 1 \\1 & 1 & 1 & -1 \\ 1 & -1 & -1 & -1 \\-1 &- 1 & 1 &- 1\\ \end{array}\right),
\frac{1}{2}\left(\begin{array}{cccc} 1 & -1 & 1 & 1 \\-1 & 1 & 1 & 1 \\- 1 & -1 & 1 & -1 \\1 &1 & 1 & -1\\ \end{array}\right) \right\rangle,
\label{autD4}
\end{equation}
\normalsize
Going back to (\ref{autD4}), it is clear that the rows of the two generators of $\mbox{Aut}(D_4)$ encode  the (entangled) eigenstates of the triple $s$ (as was the case for the generator $S$). Thus, two-qubit maximal entanglement is at the heart of the symmetries of this representation on the lattice $D_4$. The group $\mbox{Aut}(D_4)$, of order $1152$, is (up to isomorphism) the same object than the two-qubit real Clifford group $\mathcal{C}_2^+$ and the Weyl group $W(F_4)$ of the $24$-cell.

\subsection*{Three-qubit entanglement and the lattice $E_8$}

Among the several representations of the lattice $D_4$, we select the one with basis

$M=\{(2,-2,0,0,0,0,0,0),  (0,2,-2,0,0,0,0,0),  (0,0,2,-2,0,0,0,0),  (0,0,0,2,-2,0,0,0)$,

  $(0,0,0,0,2,-2,0,0),  (0,0,0,0,0,2,-2,0), (0,0,0,0,0,0,2,-2),  (1,1,1,1,1,-1,-1,-1)\}$

that has the automorphism group 

\footnotesize
\begin{eqnarray}
&\mbox{Aut}(E_8)=\left\langle g_1,g_2 \right \rangle, \nonumber \\
&\mbox{with}~~ g_1=\frac{1}{2}\left(\begin{array}{cccccccc} . & . &-1 & -1 & 1 & .&1& . \\ 1 & -1 & . & . &. &-1&.&-1 \\. & . &1 & -1 &1 & .&-1&. \\-1 &-1 &. &. &. &1& .&-1 \\ 
-1 &-1 &. &. &. &-1&.& 1\\. &. &-1 & 1 &1 & .& -1&. \\. &. &-1 & -1&-1 &.& -1& .  \\ 1 &-1 & . & . & . &1 & 0 & 1 \\    \end{array}\right) \nonumber \\
 &\mbox{and}~~ g_2=\frac{1}{2}\left(\begin{array}{cccccccc} 1 &- 1 & . & . & . & 1 &1 & . \\- 1 & 1 &. &. & . & 1&1&. \\ . & . & 1 & 1 & 1 &.&.& 1  \\ . &. &-1 &-1 & 1 & .&.&1 \\ 
 . & . & 1 & -1 & -1 &.&.& 1 \\ -1 & -1 &. &. & . & -1&1&.\\ . &. &-1 &1 & -1 &.&.&1 \\ -1 &-1 & . & . & . &1&-1& .  \\    \end{array}\right).
\end{eqnarray}
\normalsize
The group $\mbox{Aut}(E_8)$ is of order $4!6!8!=696~729~600$.

Let us consider the state encoded by the first row of $g_2$
\begin{equation}
\left|\psi\right\rangle=\frac{1}{2}(\left|000\right\rangle-\left|001\right\rangle+\left|101\right\rangle+\left|110\right\rangle).
\label{stateB}
\end{equation}
The tangles attached to the state (\ref{stateB}) are calculated, as we did for the state (\ref{state}) of the previous section. They are such that $\tau^{(3)}=\tau_{AC}=\tau_{AB}=\tau_{BC}=\frac{1}{4}$ corresponding to fully {\it balanced tripartite entanglement}. The same results holds for all states encoded by the rows of $g_1$ and $g_2$.

One can conclude that this type of entanglement, first featured in \cite{PLS2010,PlanatCPT}, is a specific property of the lattice $E_8$. 
As mentioned previously, the $3$-qubit real Clifford group $\mathcal{C}_3^+$ of order $2~580~480$ is a maximal subgroup of $\mbox{Aut}'(E_8)$ (the second largest one); the largest maximal subgroup is the automorphism group $\mbox{Aut}(E_7)$ of the lattice $E_7$.

 As a second representation of the lattice $E_8 $, we select the one obtained from the construction \lq\lq A" on the extended Hamming code $[8,4,4]$ \cite{Conway} \footnote{One may use the following commands in Magma \cite{Magma} \lq\lq  C:=HammingCode(GF(2),3); C:=ExtendCode(C);
L:=Lattice(C,"A");  
aut:=AutomorphismGroup(L:NaturalAction);aut;"}.

\footnotesize
\begin{eqnarray}
&\mbox{Aut}(E_8)=\left\langle g_1,g_2 \right \rangle, \nonumber \\
&\mbox{with}~~ g_1=\frac{1}{2}\left(\begin{array}{cccccccc} . & 1 &-1 & . & 1 & .&.& 1 \\ 1 &- 1 & . & . &1 & .&-1&. \\-1 & . &-1 & . & . & .&-1& -1 \\. &. &. &-1 &. &.& .& . \\ 
 1 &1 &. &. &-1 &.&-1.& .\\. &. &. & .&. &-1& .& . \\. &-1 &1 & .&-1 &.& .& 1  \\ 1 &. & -1 & . & . & . & 1 & -1 \\    \end{array}\right) \nonumber \\
 &\mbox{and}~~ g_2=\frac{1}{2}\left(\begin{array}{cccccccc} . & 1 & . & . & . & 1 &-1 & -1 \\ 1 & . &-1 &-1 & 1 & .&.&. \\ 1 & . & . & . &- 1 &-1&.& -1  \\ -1 &-1 &. &-1 & . & .&.&-1 \\ 
 . & . & . & 1 & 1 &.&1& -1 \\ . & -1 &-1 &1 & . & .&-1&.\\  1 & -1 &1 &. & . & 1&.&. \\ . & . & -1 & . & -1 &1&1& .  \\    \end{array}\right).
\end{eqnarray}
\normalsize

Here we get two types of generators. In the generator $g_1$, the rows $4$ and $6$ encode non entangled states while the other rows encode a state such as
\begin{equation}
\left|\psi\right\rangle=\frac{1}{2}(\left|001\right\rangle-\left|010\right\rangle+\left|100\right\rangle+\left|111\right\rangle),
\label{stateGHZ}
\end{equation}
 with $\tau^{(3)}=1$ and vanishing $2$-tangles. Such a state is of the GHZ type. The second generator $g_2$ encodes states of the type (\ref{state}) corresponding to incomplete balanced entanglement.

\subsection*{Four-qubit entanglement and the lattice $\Lambda_{16}$}

For four real qubits, the  Clifford group is $\mathcal{C}_4^+\equiv \mbox{Aut}(\Lambda_{16})$, of order $89~181~388~800$, associated to the Barnes-Wall lattice $\Lambda_{16}$ \cite{Nebe}. We use construction \lq\lq B" associated to the $[16,5,8]$ Reed-Muller code (see \cite{Conway}, p. 141) \footnote{in Magma, the code C:=ReedMullerCode(1,4);  L:=Lattice(C,\lq\lq B"); aut:=AutomophismGroup(L:NaturalAction);}. 

One gets two $16 \times 16$ gates generating $\mbox{Aut}(\Lambda_{16})$ with rows encoding entangled states such as the factorizable state
\begin{eqnarray}
&\left|\psi\right\rangle=\frac{1}{2}(\left|0000\right\rangle-\left|0011\right\rangle-\left|0101\right\rangle+\left|0110\right\rangle) \nonumber \\
&=\frac{1}{2}\left|0\right\rangle((\left|000\right\rangle-\left|011\right\rangle-\left|101\right\rangle+\left|110\right\rangle)).
\label{state3}
\end{eqnarray}
It is easy to quantify the residual tripartite entanglement of state (\ref{state3}). Using (\ref{three}), one gets true tripartite entanglement $\tau^{(3)}=1$ corresponding to the GHZ family. All states encoded by the rows in the generators of $\mbox{Aut}(\Lambda_{16})$ display vanishing $4$-tangle (the four-tangle is calculated from the expressions given in \cite{FourTangle}). Presumably, all residual three-tangles equal unity as for GHZ-type states).

\section{Entanglement arising from the $D_{12}^+$ lattice}

The unique indecomposable $12$-dimensional unimodular lattice is $D_{12}^+$. It can be generated from the construction \lq\lq A" applied to the $[12,6,6]$ extended Golay code over $GF(3)$ \footnote{In Magma, one uses the following commands \lq\lq C := GolayCode(GF(3), true);  L:=Lattice(C,"A"); aut:=AutomorphismGroup(L:NaturalAction);"}. 

The automorphism group is of order $2^{21}3^5 5^2 7^1 11$. As for the previous cases, the rows of the generators encode entangled states. If we look at the system as a bipartite $3\times 4$ system, the Schmidt rank\footnote{The Schmidt rank can easily be obtained in Magma. Magma provides the Smith normal form for any matrix $A$ with entries in an Euclidean ring (a field $K$, the ring $\mathbb{Z}$ of integers, the ring $K[x]$ of polynomials with coefficients over a field $K$...). Given an $m \times n$ matrix $A$ over an Euclidean ring $R$, one can find invertible square matrices $P$ and $Q$ such that $PAQ=S$, where $S$ is a diagonal matrix: the Smith normal form of $A$. This is equivalent to the singular value decomposition $A=P^{-1}SQ^{-1}$. The non-zero diagonal entries of the Smith form of $A$ are unique up to multiplication by a unit of $R$ and are also called elementary divisors. Thus, the Schmidt rank of $A$ equals the rank of $S$. } of the generators is found to be three. There are several types of states. We restrict our account to one of them, that we consider as the state of a $3 \times2 \times 2$ system, i.e. a qutrit coupled to two qubits.

\begin{eqnarray}
&\left|\psi\right\rangle=(2 \left|000\right\rangle+ \left|001\right\rangle-\left|010\right\rangle+\left|011\right\rangle+\left|110\right\rangle+\left|200\right\rangle)/3.
\end{eqnarray}

The reduced density matrix of the two-qubit bipartite system is found to be  

$$\rho_{BC}=\frac{1}{9}\left(\begin{array}{cccc} 5 & 2 & -2 &2 \\1 &1 & -1 & 1 \\ -2 &-1 & 2 & -1 \\2 & 1 & -1 & 1\\ \end{array}\right)$$

The set of eigenvalues of $\rho \tilde{\rho}$ associated the two-qubit subsystem $B-C$ is obtained as $\{\frac{2}{81}(10+3\sqrt{11})^{\pm 1/2},0,0\}$, corresponding the two-tangle $\tau_{BC}\sim 0.44$\footnote{
Note that the same row of the selected generator may be thought as attached to a $2\times 2\times 3$ system, i.e. two-qubits coupled to a qutrit. In that case, the state has to be rewritten as  
\begin{eqnarray}
&\left|\psi\right\rangle=(2 \left|000\right\rangle+ \left|001\right\rangle+\left|010\right\rangle-\left|100\right\rangle-\left|2000\right\rangle+\left|210\right\rangle)/3.
\label{dim12}
\end{eqnarray}
and the new reduced density matrix of the two-qubit bipartite system is found to be  

$$\rho_{BC}=\frac{1}{9}\left(\begin{array}{cccc} 2 & 2 & 1 &. \\2 &1 & 1 & . \\ 1 &1 & 2 & . \\. & . & . & .\\ \end{array}\right)$$

The set of eigenvalues of $\rho \tilde{\rho}$ associated the two-qubit subsystem $B-C$ is now obtained as $\{\frac{1}{81}(3 \pm 2\sqrt{2}),0,0\}$, corresponding the two-tangle $\tau_{BC}\sim 0.050$.  
 
Another selected generator of $\mbox{Aut}(D_{12}^+)$, seen as a $3\times 2 \times 2$ system, can easily be found to have a row with entanglement similar to that of (\ref{dim12}).}. 

Then, one can compute the reduced density matrix $\rho_{AB}$ and $\rho_{AC}$; they correspond to a qutrit/qubit system. According to the Peres criterion \cite{Horodecki}\footnote{According to the Peres partial transpose (PPT) theorem, a state $\rho_{AB}$ of a bipartite system A-B with dimension $d=d_A d_B\le 6$ is separable iff $$(\rho_{AB}^{T_A})_{i\alpha,j \beta}=(\rho_{AB}^{T_A})_{j\alpha,i \beta}$$ has non negative eigenvalues. For $d=d_A d_B> 6$, if  the partial transpose matrix $\rho_{AB}^{T_A}$ has negative eigenvalues, then the bipartite system $A-B$ is entangled. But they are states, known as bound entangled states, that satisfy the Peres positivity criterion, but still are entanged.}, a necessary and sufficient condition for the existence of entanglement between parties $A$ and $B$ is that the partial transpose matrix $\rho_{AB}^{T_A}$ contains at least one negative eigenvalue. A similar criterion holds for the entanglement between parties $A$ and $C$.

The sets of eigenvalues of matrices
$$\rho_{AB}^{T_A}=\frac{1}{9}\left(\begin{array}{cccccc} 5&.&.&2&-1&.\\.&2&2&-1&.&.\\.&2&.&.&.&.\\.&-1&.&1&.&.\\2&.&.&.&1& .\\-1&.&.&.&.&.\\ \end{array}\right),
~\rho_{AC}^{T_A}=\frac{1}{9}\left(\begin{array}{cccccc} 5&1&-1 &1&2&1\\1&2&.&.&.&.\\-1&.&1&.&.&.\\1&.&.&.&.&.\\2&.&.&.&1&.\\1&.&.&.&.&.\\ \end{array}\right).$$
are $\{0.663,0.392,0.092,0.047,-0.151,-0.0044\}$ and $\{0.717,0.211,0.111,0.047,-0.008\}$, respectively.

These calculations ensure that bipartite entanglement is spread over all parties.

\section{Entanglement arising from the Leech lattice}

The densest known lattice in dimension $24$ is the Leech lattice $\Lambda_{24}$ with kissing number $196560$. It is the unique Euclidean lattice that is even, unimodular and with non-zero vectors at least two. The automorphism group is called $\mbox{Co}_0=\mathbb{Z}_2.\mbox{Co}_1$, where the Conway group $\mbox{Co}_1$ is sporadic, with order $2^{21}.3^9.5^4.7^2.11.13.23.$

One elegant construction of $\Lambda_{24}$ makes use of the MOG (Miracle Octad Generator) coordinates, that are obtained from the codewords of the hexacode (\cite{Conway}, p. 132). In this representation, $\mbox{aut}(\Lambda_{24})$ is entangled. One can see the states encoded by the rows of the two generators as coming grom a $6 \times 4$ quantum system and for them one finds that the Schmidt rank is four. The amount of entanglement that we find with this approach is quite inhomogeneous. We restrict to a single example, the two-qubit/sextit state
\begin{eqnarray}
&\left|\psi\right\rangle=(-2 \left|001\right\rangle+ \left|100\right\rangle+\left|101\right\rangle+\left|200\right\rangle+\left|210\right\rangle+\left|300\right\rangle \nonumber \\
&+\left|311\right\rangle- \left|500\right\rangle-\left|501\right\rangle-2\left|510\right\rangle)/4,
\label{dim24}
\end{eqnarray}
with reduced density matrix
$$\rho_{BC}=\frac{1}{16}\left(\begin{array}{cccc} 4 & 2 & 3 &1 \\2 &6 & 2 & . \\ 3 &2 & 5 & . \\1 & . & . & 1\\ \end{array}\right)$$
and eigenvalues of $\rho \tilde{\rho}$ as $\{\frac{1}{64}(9\pm 4 \sqrt{2}),\frac{1}{256}(3\pm 2 \sqrt{2})\}$. It follows that the two-tangle for this states is $\tau_{BC}\sim 0.00536$.

It can be shown that the partial traces $\rho_{AB}^{T_A}$ and $\rho_{AC}^{T_A}$ violate the Peres positivity criterion so that there also is entanglement between the parties $A-B$ and $A-C$.

It would be desirable to analyze the entanglement in $\Lambda_{24}$ as arising from a three-qubit/qutrit system but the methods to achieve this goal seem not yet to be available.

\section{Conclusion}   

The symmetry group acting on the generating basis of even Euclidean lattices has been described as a group of entangling quantum gates. Barnes-Wall lattices are known to be related to the Clifford group of quantum error correction \cite{Nebe,Calderbank} and to discretised Hilbert spaces \cite{Hopf} but the occurrence of Leech lattice is novel in this context. It may be that real world systems, like quark systems, or codons, or even black holes \cite{Levay}, exploit dimension $24=2^3 \times 3 $ as three-qubits and qutrits at the same time.  Leech lattice may also be constructed from Leech roots in Lorentzian space $\mathbb{R}^{25,1}$ (see \cite{Conway}, chaps. 26-27) and dimension $26$ corresponds to the original version of string theory through the no-ghost theorem. These new vistas (see also \cite{Allcock}), and their connection to tripartite entanglement,  will be explored in future work.

\section*{Acknowledgements} The author acknowledges Bernd Souvignier for his reading of an earlier version of the manuscript and R. Mosseri for his feedback about the geometry of the discretised Hilbert space. The work started in December 2009 during a visit at de Br\'un Center for Computational Algebra in Galway under the invitation of Michael Mc Gettrick. It was pesented at International Conference on Quantum Optics and Quantum Computing (ICQOQC-11) in Noida and at the 18th Central European Workshop on Quantum Optics (CEWQO 2011) in Madrid.

\end{document}